\documentclass[%
 reprint,
 amsmath,amssymb,
 aps]{revtex4-2}

\usepackage{graphicx}
\usepackage{dcolumn}
\usepackage{xcolor}
\usepackage{bm}
\usepackage{hyperref}
\usepackage[mathlines]{lineno}

\begin{document}

\preprint{APS/123-QED}

\title{Aging in binary-state models: The Threshold model for Complex Contagion}

\author{David Abella}\email{david@ifisc.uib-csic.es}%
\author{Maxi San Miguel}%
\author{Jos\'e J. Ramasco}
\affiliation{%
 Instituto de F\'{\i}­sica Interdisciplinar y Sistemas Complejos IFISC (CSIC - UIB), Campus Universitat Illes Balears, 07122 Palma de Mallorca, Spain
}%

\date{\today}%

\begin{abstract}
Binary-state models are those in which the constituent elements can only appear in two possible configurations. These models are fundamental in the mathematical treatment of a number of phenomena such as spin interactions in magnetism, opinion dynamics, rumor and information spreading in social systems, etc. Here, we focus on the study of non-Markovian effects associated with aging for binary-state dynamics in complex networks. Aging is considered as the property of the agents to be less prone to change state the longer they have been in the current state, which gives rise to heterogeneous activity patterns. We analyze in this context the Threshold model of Complex Contagion, which has been proposed to  explain, for instance, processes of adoption of new technologies and in which the agents need the reiterated confirmation of several contacts (until reaching over a given neighbor fraction threshold) to change state. Our analytical approximations give a good description of extensive numerical simulations in Erd\"os-R\'enyi, random-regular and Barab\'asi-Albert networks. While aging does not modify the spreading condition, it slows down the cascade dynamics towards the full-adoption state: the exponential increase of adopters in time from the original model is replaced by a stretched exponential or power-law, depending on the aging mechanism. Under several approximations, we give analytical expressions for the cascade condition and for the exponents of the exponential, power-law and stretched exponential growth laws for the adopters density. Beyond networks, we also describe by numerical simulations the effects of aging for the Threshold model in a two-dimensional lattice. 
\end{abstract}

\maketitle


\section{\label{sec:Introduction} Introduction}

Stochastic binary-state models are a versatile tool to describe a variety of natural and social phenomena in systems formed by many interacting agents. Each agent is considered to be in one of two possible states: susceptible/infected, adopters/non-adopters, democrat/republican, etc, depending on the context of the model. The interaction among agents is determined by the underlying network and the dynamical rules of the model. There are many examples of binary-state models, including processes of opinion formation \cite{Voter-original,sood-2005,fernandez-gracia-2014,redner-2019}, disease or social contagion \cite{granovetter-1978,pastor-satorras-2015}, etc. Extended and modified versions of these models can lead to very different dynamical behaviors than in the original model. As examples, the use of multi-layer  \cite{diakonova-2014,diakonova-2016,amato-2017} or time dependent\cite{vazquez-2008} networks, higher-order interactions \cite{de-arruda-2020, iacopini-2019, cencetti-2021}, non-linear collective phenomena \cite{castellano-2009,peralta-2018}, noise \cite{carro-2016} and non-Markovian \cite{van-mieghem-2013,starnini-2017,peralta-2020A,chen-2020} effects induce significant changes to the dynamics.

A well known binary-state model is the Threshold model \cite{watts-2002}, introduced by Mark Granovetter \cite{granovetter-1978}, for rumor propagation, adoption of new technologies, riots, stock market herds, political and environmental campaigns, etc. These are examples of {\it Complex Contagion} processes \cite{centola-2007,unknown-author-2018} in which contagion, at variance with {\it Simple Contagion} (such as 
in the Voter and  SIS models) requires simultaneous exposure to multiple adopter neighbors and a threshold fraction of neighboring agents that have already undergone contagion. Complex contagion implies a nonlinear process of group or many-agent interactions built from a combination of pairwise interactions. The discontinuous phase transition and the cascade condition exhibited by the Threshold model were predicted with analytical tools in Ref. \cite{watts-2002}. This model has been extensively studied in regular lattices and small-world networks \cite{centola-2007}, random graphs \cite{gleeson-2007},  modular and community structure \cite{gleeson-2008}, clustered networks \cite{hackett-2011,hackett-2013}, hypergraphs \cite{de-arruda-2020}, homophilic networks \cite{diaz-diaz-2022}, etc. Moreover, recent studies also include variants of the adoption rules including
the impact of opinion leaders \cite{liu-2018} and seed-size \cite{singh-2013}, on-off threshold \cite{dodds-2013} and the competition between simple and complex contagion \cite{czaplicka-2016,min-2018,diaz-diaz-2022}. Additionally, the Threshold model has been confronted with several sources of empirical data \cite{Centola-2010,karimi-2013,karsai-2014,rosenthal-2015,karsai-2016,mnsted-2017,unicomb-2018,guilbeault-2021}.

Theoretical and computational studies of stochastic binary-state models, including the Threshold model, usually rely on a Markovian assumption for its dynamics. 
However, there is strong empirical evidence against this assumption in human interactions.  For example, bursty non-Markovian dynamics with heavy-tail inter-event time distributions, reflecting temporal activity patterns,  have been reported in many studies \cite{iribarren-2009,karsai-2011,rybski-2012,zignani-2016,artime-2017,kumar-2020}. The understanding of these non-Markovian effects is in general a topic of current interest \cite{van-mieghem-2013,starnini-2017,peralta-2020C,peralta-2020A}. In particular, for the Threshold model, memory effects have been included as past exposures memory \cite{dodds-2004}, message-passing algorithms \cite{shrestha-2014}, memory distributions for retweeting algorithms \cite{gleeson-2016} and timers \cite{oh-2018}.

Aging is an important non-Markovian effect that we address in this paper for binary-state models. Aging accounts for the influence that the persistence time of an agent in a given state modifies the transition rate to a different state \cite{stark-2008,fernandez-gracia-2011,perez-2016,boguna-2014,chen-2020}, so that, the longer an agent remains in a given state, the smaller is the probability to change it. Aging effects have been already shown to modify binary-state dynamics very significantly. For example, aging is able to produce coarsening towards a consensus state in the Voter model \cite{fernandez-gracia-2011,peralta-2020C}, to induce a continuous phase transitions in the noisy Voter model \cite{artime-2018,peralta-2020A} or to modify qualitatively the phase diagram and non-equilibrium dynamics of Schelling segregation model \cite{Abella}.


In this paper, we provide a general theoretical framework to discuss aging effects building upon a general Markovian approach for binary-state models \cite{gleeson-2011,gleeson-2013} and we apply it to the Threshold model of Complex Contagion. We build a general master equation for any binary-state model with temporal activity patterns and we propose two different aging mechanisms giving rise to heterogeneous activity patterns. 
Theoretical predictions are matched with extensive numerical simulations in different networks. In addition, the role of both aging mechanisms is also studied in a regular two-dimensional lattice. 

The paper is organized as follows. In the next section, we describe the original Threshold model and introduce exogenous and endogenous aging in the model. In section \ref{sec:Complex networks}, numerical results are reported and contrasted with theoretical predictions for different complex networks. For completeness, in section \ref{sec:Lattice} the case of a 2D-lattice is analyzed. The final section contains a summary and a discussion of the results. The derivation of the general Master Equation for binary-state dynamics with aging effects is given in the Appendix.

\section{\label{sec:Threshold model with aging} Aging and the Threshold model}

In the standard Threshold model \cite{granovetter-1978,watts-2002}, one considers a network of $N$ interacting agents. Each node of the network represents an agent $i$ with a binary-state variable $\sigma_i = [0,1]$  and a given threshold $T$ ($0<T<1$). The state indicates if the agent has adopted a technology (or joined a riot, spread a meme or fake-new, etc) or not. We use the wording of a technology adoption process for the rest of the paper. If a node $i$ (with $k$ neighbors) has not adopted  ($\sigma_i = 0$) the technology, becomes adopter ($\sigma_i = 1$) if the fraction $m / k$ of neighbors adopters exceeds the threshold $T$. Adopter nodes cannot go back to the non-adopter state.

In the Threshold model with aging, each agent has an internal time $j = 0,1,2,...$  (in Monte-Carlo units) as in Refs. \cite{fernandez-gracia-2013,artime-2018,peralta-2020C,peralta-2020A,chen-2020,fernandez-gracia-2011,perez-2016,stark-2008,Abella}.  As initial condition, we set $j = 0$ for all nodes. In numerical simulations, we follow a Random Asynchronous Update in which agents are activated with a probability $p_{A} (j) = 1/(j+2)$. When a non-adopter agent is activated, she changes state according to the threshold condition $m/k > T$. We will consider two different aging mechanisms, endogenous and exogenous aging \cite{fernandez-gracia-2011}, which account for the power-law inter-event time distributions empirically observed in human interactions \cite{artime-2017}. In the endogenous aging the internal time measures the time spent in the current state: If an agent in an updating attempt is not activated or does not adopt, the internal time increases one unit. Therefore, the longer an agent has remained without adopting the technology, the more difficult it is for her to adopt it. 

In the exogenous aging, the internal time accounts for the time since the last attempt to change state: In each updating attempt in which the agent is activated, the internal clock resets to $j = 0$ even if there is adoption. In this case, aging is understood as a resistance to adopt the technology the longer the agent has not been induced to consider adoption by some external influence.  


\section{\label{sec:Complex networks} Complex networks}

\begin{figure}[]
\includegraphics[width=1\columnwidth]{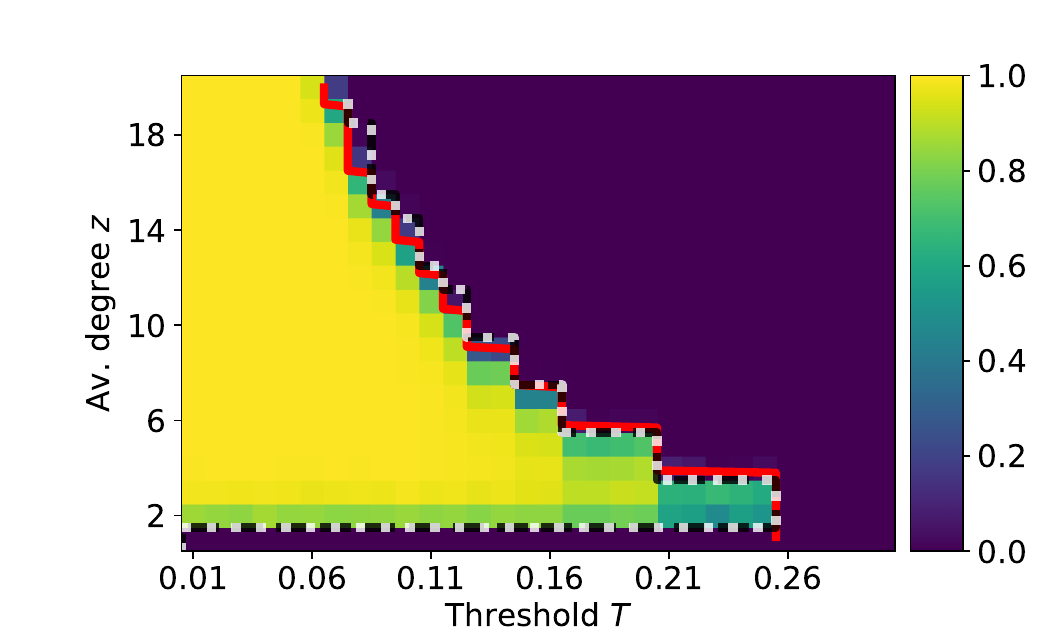}
\caption{\label{fig:umbral} Average density $\rho$ of adopters for an Erd\"os-R\'enyi graph of mean degree $z$ using a model with threshold $T$. Color-coded values of $\rho$ are from numerical simulations of the model without aging in a graph with $N = 10000$ agents. Black dashed and white doted lines correspond to the critical threshold value obtained numerically for the model with exogenous and endogenous aging, respectively. The red solid line is the analytical approximation of cascade boundary, from Eq. \ref{eq:linear}, which is the same with and without aging.}
\end{figure}

In this section we discuss the Threshold model with endogenous and exogenous aging in three different complex networks: random-regular \cite{wormald1999models}, Erd\"os-R\'enyi \cite{erdos1960evolution} and Barab\'asi-Albert \cite{barabasi2009scale}.

\subsection*{\label{subsec:Numerical results} Numerical results}

\begin{figure}
\includegraphics[width=1\columnwidth]{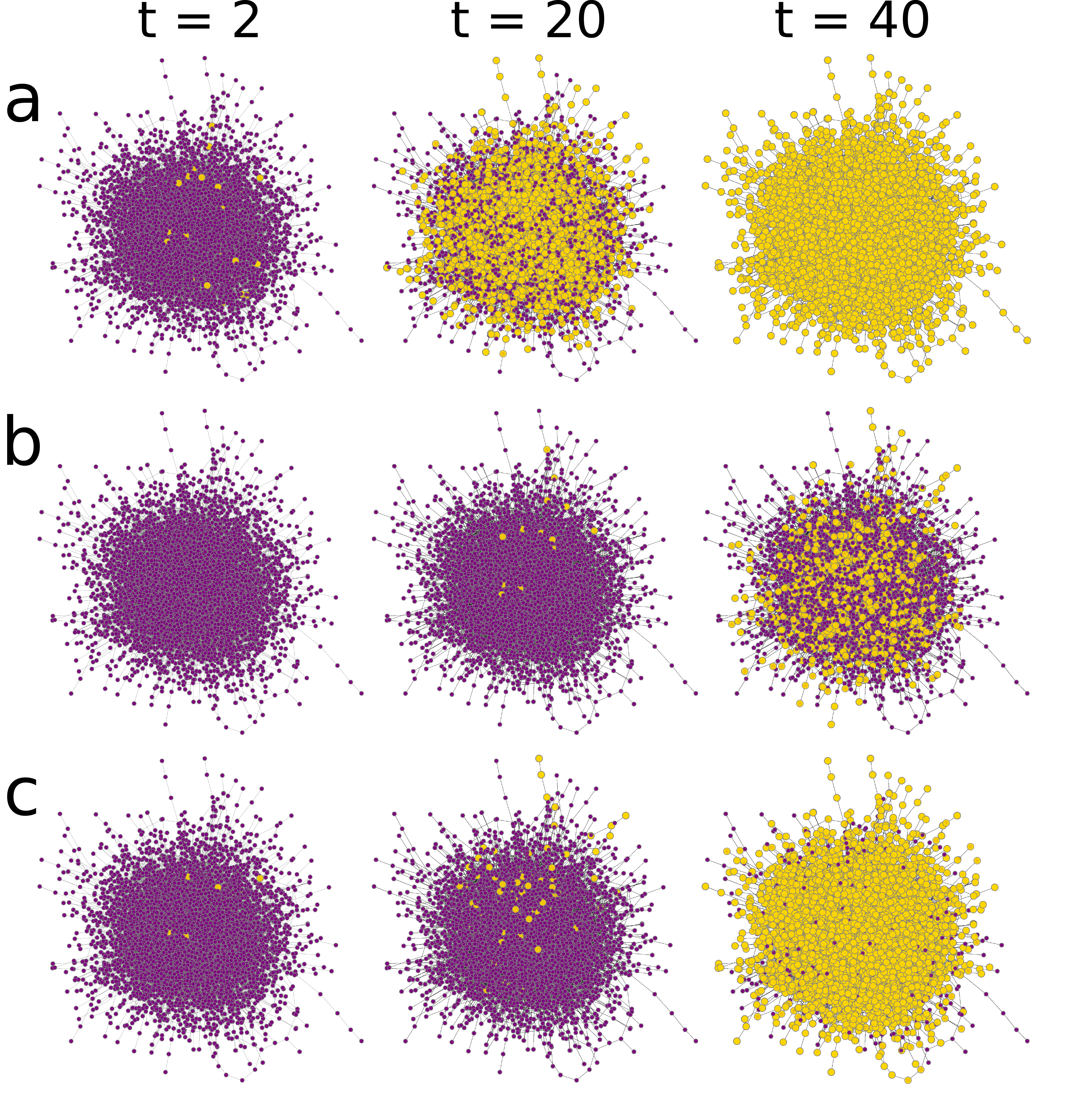}
\caption{\label{fig:graph_plot} Cascade spreading for the original Threshold model (a), and the versions with endogenous (b) and exogenous (c) aging. Yellow nodes have adopted and purple nodes not. Time increases from left to right. Simulations are performed in an Erd\"os-R\'enyi network with degree $z = 3$ and $T = 0.22$. System size is $N = 8000$.}
\end{figure}

\begin{figure}
\includegraphics[width=\columnwidth]{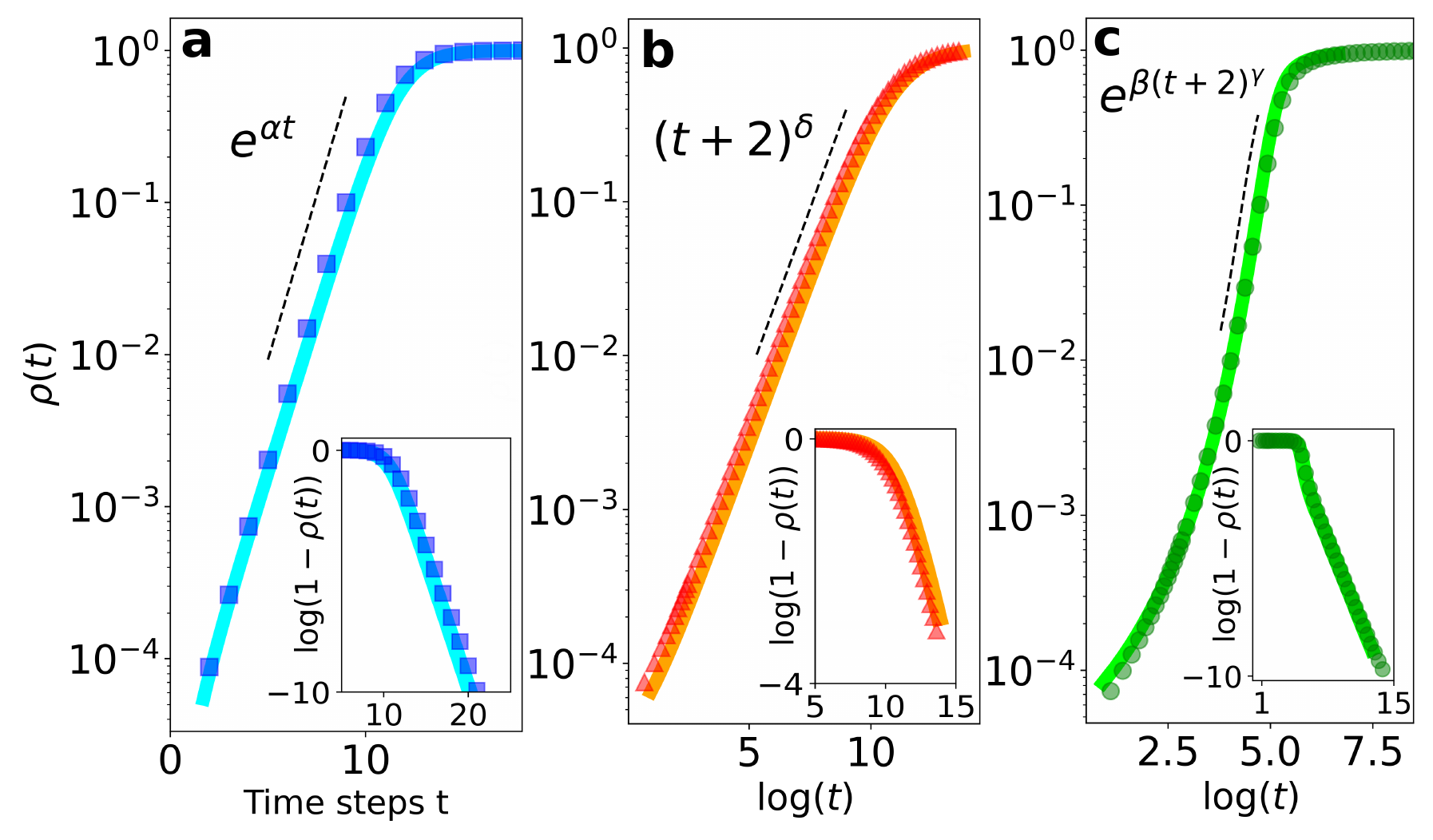}
\caption{\label{fig:models} Cascade dynamics and fall to the full-adopt state ($\rho = 1$) of the Threshold model without aging (a) and the versions with endogenous (b) and exogenous (c) aging effects. At (b-c), the evolution is plotted as a function of the logarithm of time $\log{(t)}$ in Monte Carlo steps. The underlying network is a 3-regular random graph and the homogeneous threshold is $T = 0.2$. The exponent values are $\alpha \simeq 1.0$, $\beta \simeq 1.14$, $\gamma \simeq 0.38$ and $\delta \simeq 1.0$. Numerically integrated solutions of Eq.\ref{eq:AME_Threshold} (solid lines) describe accurately the numerical results.}
\end{figure}

For the networks considered, the Threshold model undergoes a discontinuous phase transition at a certain critical value $T_{c}$ \cite{watts-2002}. For $T<T_c$, a small initial seed of adopters triggers a global cascade where all agents in the system adopt the technology (change from $\sigma_i = 0 \mbox{ to } \; 1$). In our analysis, the initial condition is set to favor cascades: one random agent $i$ with degree $k_i = z$ and all her neighbors are initially adopters, as in Ref. \cite{centola-2007,singh-2013}. For $T>T_c$, there are few cascade occurrences and none of them is global. The critical threshold dependence with the average degree $z$ of the underlying network has been studied in Refs. \cite{watts-2002, gleeson-2007}. For the two aging mechanisms considered, numerical simulations show that the critical threshold $T_c$ dependence on $z$ is very similar to the one for the model without aging (see Fig. \ref{fig:umbral}). Therefore, for large connected networks, tends to the same cascade condition derived for the original Threshold model $T_{c} = 1 / z$ \cite{watts-2002}.
Threshold model does not modify the critical values of the threshold parameter, it has a large impact in the dynamics of the cascade process of complex contagion (Fig.\ref{fig:graph_plot}). 
From numerical simulations we find that, without aging,  the average fraction of adopters follows an initial exponential increase with time (see Fig. \ref{fig:models}a), 
\begin{equation}
\rho(t) \sim \rho_0 \, e^{\alpha \, t},
\end{equation}
where $\rho_0$ is the initial fraction of adopters (seed). This behavior is universal for all values of the control parameters $z$ and $T$ below the cascade condition. The dependence of the exponent with these parameters $\alpha (z,T)$ is shown in Fig. \ref{fig:endo_exp}. In addition, we investigated the approach to the full-adopt state ($\rho = 1$) and we found that the number of non-adopters follows an exponential decay $1 - \rho(t) \sim e^{-t}$ for all values of the control parameters (see inset in Fig.\ref{fig:models}a).

When aging is introduced, the cascade dynamics are much slower than an exponential law. 
For endogenous aging, all agents non-adopters have the same activation probability $p_A(j)$, which decreases at each time step. This gives rise to a cascade dynamics well-fitted by a power law initial increase (see Fig. \ref{fig:models}b),
\begin{equation}
\rho(t) \sim \rho_0 \, ((t + 2) / 2)^\delta .
\end{equation}
For exogenous aging, we observe a slow adoption spread at the beginning followed by a cascade where almost all agents adopt the technology (Fig. \ref{fig:graph_plot}b). This behavior is well-fitted with a stretched exponential increase of the number of adopters (see Fig. \ref{fig:models}c),
\begin{equation}
\rho(t) \sim  \rho_0\,  e^{\beta \, ((t + 2) / 2)^{\gamma}} .
\end{equation}
For both aging mechanisms, in the last stages of evolution, a few ``stubborn'' non-adopters remain, although the environment favours the adoption. Due to the chosen activation probability, the number of non-adopters decay with a power law $1 - \rho(t) \sim 1/(t+2)$ in both cases (see inset at Fig. \ref{fig:models}(b-c)).

\begin{figure*}
\includegraphics[width=\linewidth]{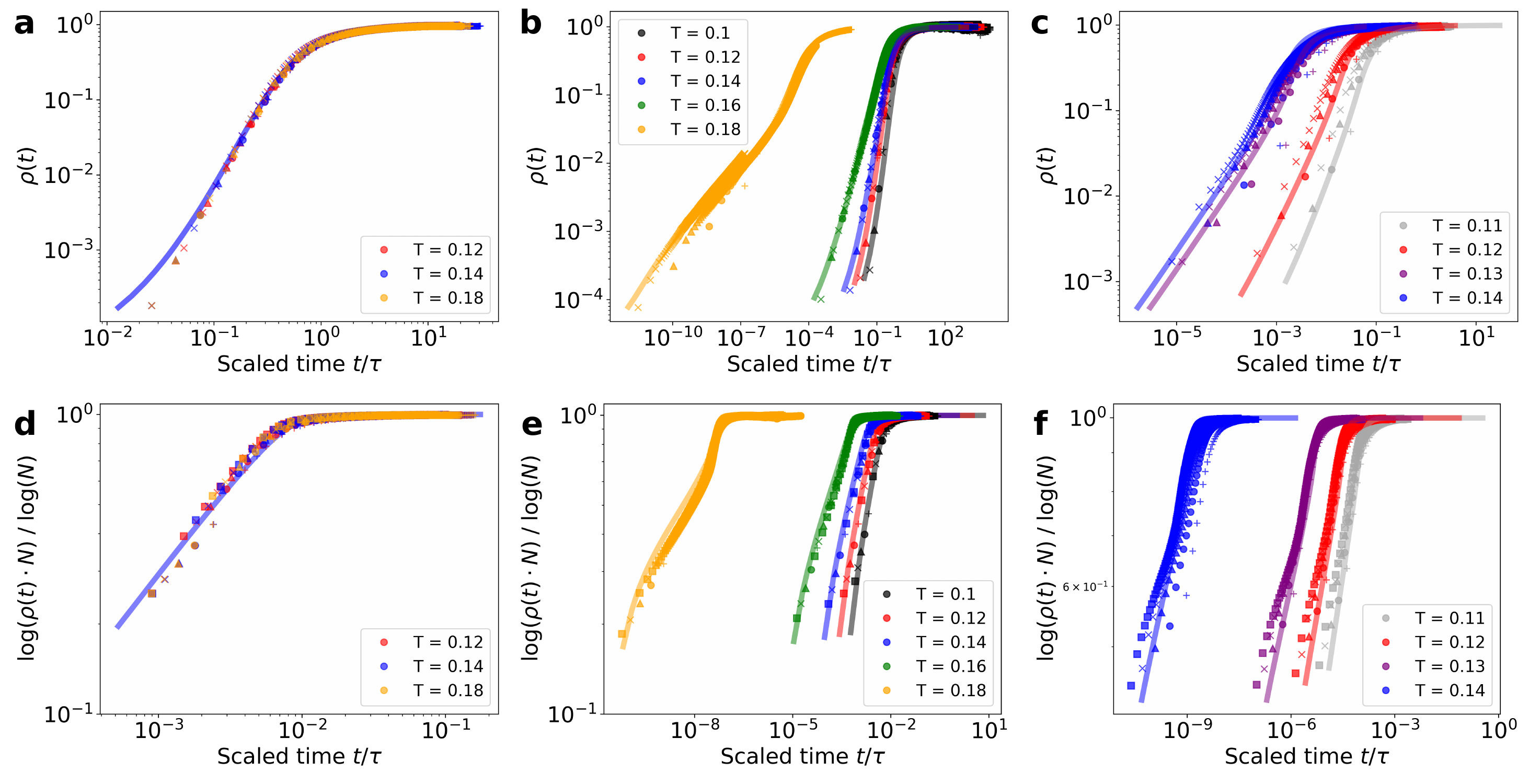}
\caption{\label{fig:exo_endo_evo} Cascade dynamics of the Threshold model with endogenous (a - c) and exogenous (d - f) aging. From the left column to the right: a regular random graph with degree $z=5$ (a and d), an Erd\"os-R\'enyi Graph with average degree $z = 5$ (b and e) and a Barab\'asi-Albert Graph with average degree $z = 8$ (c and f). Different colors indicate different values of $T$ and markers correspond to different system sizes: $N = 2500$ (plus), $10000$ (circles), $40000$ (triangles), $160000$ (crosses) and $640000$ (squares). Time is scaled according to the system size for each model: $\tau_{\rm{EXO}} = 2(\log(N)/\beta)^{1/\gamma} - 2$, $\tau_{\rm{ENDO}} = 2N^{1/\delta} - 2$. Solid lines are obtained from the solutions of Eq. \eqref{eq:PA}.}
\end{figure*}

The power law and the stretched exponential dynamics for endogenous and exogenous aging, respectively, are observed for all parameter values $z$ and $T$ below the cascade condition ($T < T_c$) and for all system sizes. This is shown in Fig. \ref{fig:exo_endo_evo} for a random regular, Erd\"os-R\'enyi and  Barab\'asi-Albert networks. In particular, we show that the time dependence for different system sizes collapse in a single curve when time is scaled with the system size dependent timescale that follows from either the power law dynamics $(\tau_{\rm{ENDO}} = 2N^{1/\delta} - 2)$  or the stretched exponential law  $(\tau_{\rm{EXO}} = 2( \log(N)/\beta )^{1/\gamma} - 2)$. Notice that the scaling of the y-axis is necessary in Fig.\ref{fig:exo_endo_evo}(d-f) to recover a linear dependence (for all system sizes) due to the stretched exponential increase.


A different question is the dependence of the exponents of the power law and stretched exponential with the parameters $z$ and $T$. Numerical results for $\delta(z,T)$ and $\gamma(z,T)$ are shown in Figs. \ref{fig:endo_exp} and \ref{fig:exo_exp}. For a random-regular graph, as apparent from Fig. \ref{fig:exo_endo_evo}, the exponents do not depend on the parameter $T$ up to $T_c$ (so the exponents are dependent only on $z$, $\gamma(z)$ and $\delta(z)$), while for Erd\"os-R\'enyi and Barab\'asi-Albert networks
$\gamma(z,T)$ and $\delta(z,T)$ decrease with $T$ when approaching $T_c$, indicating a slowing down of the dynamics. Also, for these two latter networks, the exponents present a maximum value at a certain value of $z$. This maximum value at a certain $z$ for a fixed $T$ can be understood as being between the two critical lines of Fig. 1.

\subsection*{\label{subsec:Approximate master equation and solutions} General mathematical description and differential equations}

To account for the non-Markovian dynamics introduced by the aging mechanism, we need to go beyond the standard mathematical descriptions of the Threshold model \cite{gleeson-2007,gleeson-2008,gleeson-2013}. We do so using a Markovian description by enlarging the number of variables \cite{peralta-2020C,peralta-2020A}. Namely, we classify the agents with degree $k$, number of adopter neighbors $m$ and age $j$ as different sets in a compartmental model in a general framework for binary-state dynamics in complex networks \cite{watts-2002, gleeson-2011,gleeson-2013}. Assuming a local tree-like network structure, as the one generated using the configuration model for a generic degree distribution $p_k$ \cite{molloy-1995,newman-2001} or Erd\"os-R\'enyi model, we derive a general master equation \footnote{We use here the term  ``master equation'' for consistency with  Refs. \cite{gleeson-2011,gleeson-2013}, but the word ``master'' has a different meaning than the one used to describe an equation for the probability distribution \cite{peralta-2020B}} for binary-state dynamics with temporal activity patterns in complex networks considering the following possible transitions (see Appendix \ref{app:DERIVATION OF MASTER EQUATION WITH AGING} for details):
\begin{itemize}
    \item A susceptible (infected) node changes state and resets internal age with probability $F (k,m,j)$ ($R (k,m,j)$);
    \item A susceptible (infected) node remains in the same state and resets internal age to zero ($j \to 0$) with probability $F_R (k,m,j)$ ($R_R (k,m,j)$);
    \item A susceptible (infected) node remains in the same state and ages ($j \to j+1$) with probability $F_{A} (k,m,j)$ ($R_{A} (k,m,j)$).
\end{itemize}
See an schematic representation in Fig. \ref{fig:ame_plot1}. Note that we are using here epidemics notation of susceptible/infected nodes \cite{gleeson-2011,gleeson-2013}, but it is immediately translated to the non-adopter/adopter situation of our model. For the specific case of the Threshold model, dynamics are monotonic and $R (k,m,j) = 0$. Moreover, when an agent becomes an adopter, there are neither resetting nor aging events $R_R (k,m,j) = R_A (k,m,j) = 0$. This means as well that equations for the susceptible and infected nodes are independent. Thus, we can write the following rate equations for the evolution of the fraction $s_{k,m,j} (t)$ of $k$-degree susceptible nodes with $m$ infected neighbors and age $j$:
\begin{align}
\label{eq:AME_Threshold}
   \frac{d s_{k,m,j}}{dt} = & \,  - s_{k,m,j} - (k-m)\, \beta^s \, s_{k,m,j} \nonumber\\ 
   & + (k-m+1) \, \beta^s \, s_{k,m-1,j-1}  \nonumber\\
   & + F_{A} (k,m,j-1)\, s_{k,m,j-1},  \\
     \frac{d s_{k,m,0}}{dt}  = & \,   - s_{k,m,0} - (k - m)\, \beta^s   \,s_{k,m,0} \nonumber \\
    & + \sum_{l = 0} F_{R} (k,m,l)\, s_{k,m,l}, \nonumber 
\end{align}
where $\beta^s$ is a non-linear function of $s_{k',m',j'}$ for all values of $k'$,$m'$ and $j'$  (see Eq. \eqref{beta_s}). The remaining step is to define explicitly the transitions probabilities for our aging mechanisms. For both exogenous and endogenous aging, the infection probability is the probability that the node is activated and has a fraction of adopters that exceeds the threshold $T$, which means that 
\begin{equation}
F(k,m,j) = p_A(j) \, \theta(m/k - T),
\end{equation} 
where $\theta()$ is the Heaviside step function. 

The reset and aging probabilities for endogenous and exogenous aging mechanisms are different. The simplest case is the endogenous aging where there is no reset $F_{R} (k,m,j) = 0$ and agents age with probability 
\begin{align}
F_{A} (k,m,j) =& \,  1 - F(k,m,j) \\ = & \, 1 - p_{A}(j)\, \theta \left(m/k - T\right). \nonumber
\end{align}
When aging is exogenous, the reset probability is the probability to activate and not adopt 
\begin{equation}
F_R (k,m,j) = p_A (j)\, \left(1 - \theta \left(m/k - T\right)\right). 
\end{equation}
Thus, agents that age are just the ones that do not activate, $F_A (k,m,j) = 1 - p_A(j)$.

Using these definitions, we have solved Eq. \eqref{eq:AME_Threshold} for the Threshold model with both endogenous and exogenous aging. Solutions give  good agreement with numerical simulations (see Fig. \ref{fig:models}). However, in a general network, considering a cutoff for the degree $k = 0,\dots,k_{\rm{max}}$ and age $j = 0,\dots,j_{\rm{max}}$, the number of differential equations to solve is $(k_{\rm{max}} + 1)\, (k_{\rm{max}} + 1)\, (j_{\rm{max}} + 1)$, which grows fast with the largest degree and largest age considered. Therefore, some further approximations are needed to obtain a convenient reduced system of differential equations. 

As an ansatz, we assume that timing interactions can be effectively decoupled from the adoption process, so that the solution of Eq. \eqref{eq:AME_Threshold} can be written as
\begin{equation}
    \label{eq:assumption1}
    s_{k,m,j}(t) = s_{k,m}(t) \, G_{j} (t),
\end{equation}
where $s_{k,m}$ is the fraction of susceptible nodes with degree $k$ and $m$ infected neighbors $s_{k,m} = \sum_{j} s_{k,m,j}$ and there is an age distribution $G_{j} (t)$, independent of the adoption process.

If we sum over the variable age $j$ in Eq. \eqref{eq:AME_Threshold}, we can rewrite the following rate equations for the variables $s_{k,m}$
\begin{align}
    \label{eq:threshold_AME_red}
    \frac{d s_{k,m}}{dt}  = & \,  - \langle p_A \rangle \, \theta(m - kT)\, s_{k,m} \\
    & - (k - m) \, \beta^s \,  s_{k,m} + (k - m + 1)\, \beta^s \,  s_{k,m-1}, \nonumber
\end{align}
where aging effects are  just included in $\langle p_A \rangle(t)$: 
\begin{equation}
    \langle p_A \rangle(t) = \sum_{j = 0} p_A(j) \, G_j (t).
\end{equation}

Using the definition of the fraction of k-degree infected agents $\rho_k (t)$,
\begin{equation}
    \rho(t) = 1 - \sum_j \sum_{m = 0}^k s_{k,m,j},
\end{equation}

and along lines of Ref. \cite{gleeson-2013}, we use the exact solution

\begin{equation}
    s_{k,m} = (1 - \rho_k (0)) \, B_{k,m}[\phi],
\end{equation}

where $B_{k.m}[\phi]$ is the binomial distribution with $k$ attempts, $m$ successes and with success probability $\phi$. From this point, we derive from Eq. \eqref{eq:threshold_AME_red} a reduced system of two coupled differential equations for the fraction of adopters $\rho(t) = \sum_k p_k \rho_k (t)$ and an auxiliary variable $\phi (t)$ (see details in Ref. \cite{gleeson-2013}):
\begin{align}
    \label{eq:PA}
        & \frac{d \rho}{dt} = \langle p_A \rangle [ h(\phi) - \rho ], \nonumber\\
        \\
        & \frac{d \phi}{dt} = \langle p_A \rangle [ g(\phi) - \phi ], \nonumber
\end{align}
where $\phi(t)$ can be understood as the probability that a randomly chosen neighbor of a susceptible node is infected at time $t$. The functions $h(\phi)$ and $g(\phi)$ are nonlinear functions of this variable $\phi$
\begin{align}
    h (\phi)  = & \,  \sum_k p_k\,  \left( \rho_k (0) + (1 - \rho_k (0))\,  \sum_{m = kT}^{k} B_{k,m}[\phi]\right),\nonumber\\
    \\
    g (\phi)  = & \, \sum_k \frac{k}{z}\,  p_k \,  \left( \rho_k (0) + (1 - \rho_k (0)) \, \sum_{m = kT}^{k} B_{k-1,m}[\phi]\right). \nonumber
\end{align}
 When $\langle p_A \rangle$ is replaced by a time-independent constant, Eqs. \eqref{eq:PA} reduce to previous results for the original model \cite{gleeson-2008}.
 
Determining the distribution $G_j (t)$ is not a priory simple. For endogenous aging, all non-adopters have the same age at each time step and $G_j (t) = \delta(j-t)$. Therefore, $\langle p_A \rangle = 1/(t+2)$. The numerical solution of Eq. \eqref{eq:PA} gives a good agreement with numerical simulations (see Fig. \ref{fig:exo_endo_evo}(a-c)). For the case of exogenous aging, the reset of the internal clock makes more difficult a choice for $G_j (t)$.  Inspired on the stretched exponential behavior of $\rho(t)$ observed from numerical simulations, we propose $\langle p_A \rangle = 1/(t+2)^\mu$. For $\mu = 0.75$, the numerical integration of Eq. \eqref{eq:PA} gives a very good agreement with our simulations (see Fig. \ref{fig:exo_endo_evo} (d-f)).

\begin{figure}
    \includegraphics[width=\columnwidth]{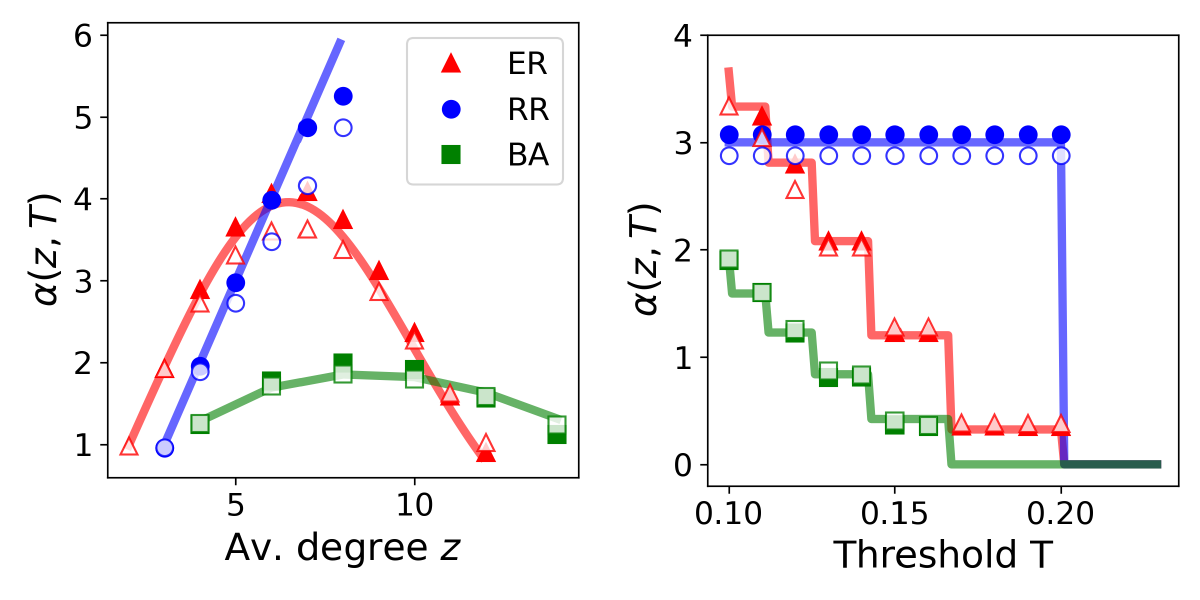}
    \caption{\label{fig:endo_exp}  Exponent $\alpha$ for the original Threshold model (filled markers) and $\delta$ for the version with endogenous aging (empty markers) for different values of the average degree $z$ (and $T = 0.1$) (left) and as a function of $T$ for fixed $z$ (right). Different markers indicate results from numerical simulations with different topology: red triangles indicates an Erd\"os-R\'enyi (ER) Graph, blue circles indicate a Random Regular (RR) Graph and green squares indicate a Barab\'asi-Albert (BA) graph. In the right panel, the average degree is fixed $z = 5$ for ER and RR, and $z = 8$ for the BA. Predicted values by Eq. \eqref{eq:alpha} (solid lines) fit the results for each topology. System size is fixed at $N = 4 \times 10^6$ for the original model and $N = 3.2 \times 10^5$ for the version with aging.}
\end{figure}

\subsection*{\label{subsec Analytical results} Analytical results}

To obtain an analytical result for the cascade condition and for the exponents of the predicted exponential, stretched-exponential and power-law cascade dynamics that we fitted from numerical simulations, we need to go a step beyond the numerical solution of our approximated differential equations (Eqs. \eqref{eq:AME_Threshold} and \eqref{eq:PA}). 

For a global cascade to occur, it is needed that the variable $\phi(t)$ grows with time. If we assume a small initial seed ($\rho_k (0) \; \to \; 0$), Eq. \eqref{eq:PA} can be rewritten as \cite{gleeson-2007}
\begin{equation} 
    \label{eq:pre_lin}
    \frac{d \phi}{dt}  = \langle p_A \rangle \, \left( -\phi + \sum_{k=1}^{\infty} \frac{k}{z} \, p_k \, \sum_{m = k\, T}^{k} B_{k-1,m} [\phi] \right).
\end{equation}
Rewriting the sum term as $\sum_{l=0}^{\infty} C_l \, \phi^l$, with coefficients 
\begin{equation}
    \label{eq:coef_phi}
    C_l = \sum_{k=l}^{\infty} \sum_{m=0}^{l} { k-1 \choose l} \, {l \choose m} \, (-1)^{l+m} \, \frac{k}{z} \, p_k \, \theta\left(m/k - T \right),
\end{equation}
we linearize Eq. \eqref{eq:pre_lin} around $\phi = 0$:
\begin{equation}
    \label{eq:linear}
    \frac{d \phi}{dt} \approx  \langle p_A \rangle \, ( C_1 -1) \, \phi.
\end{equation}
The solution for Eq. \eqref{eq:linear} is then
\begin{equation}
    \label{eq:phi_general}
    \phi(t) = \rho_0\,  e^{(C_1 - 1) \, \int_0^t \langle p_A \rangle (s) \, ds},
\end{equation}
given that $ \phi(0) = \rho_0$.

Since $\langle p_A \rangle(t)$ is always positive, global cascades occur when $( C_1 - 1) > 0 $. This cascade condition does not depend on the aging term $\langle p_A \rangle(t)$ and, thus, it is the same as for the Threshold model without aging. In Fig. \ref{fig:umbral}, the red solid line is the result of this analytical calculation, and it is in good agreement with the numerical results. 

\begin{figure}
\includegraphics[width=\columnwidth]{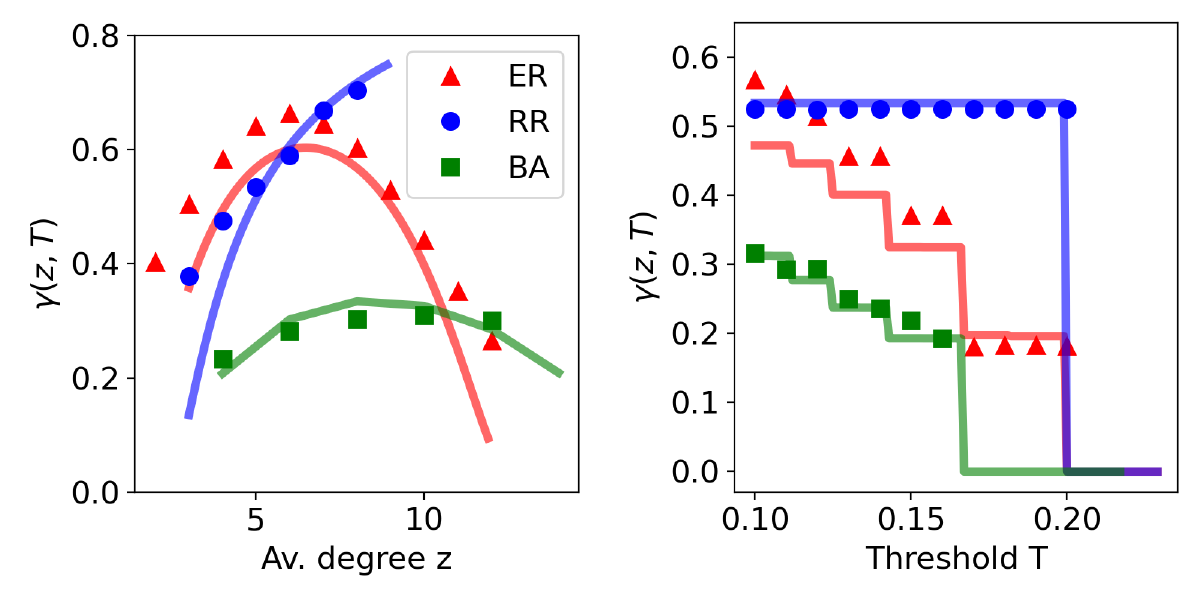}
\caption{\label{fig:exo_exp} Exponent $\gamma$ for the Threshold model with exogenous aging for different values of the average degree $z$ ($T = 0.1$) (left) and as a function of  $T$ for fixed $z$ (right). Different markers indicate results from numerical simulations with different topology: red triangles indicates an Erd\"os-R\'enyi (ER) Graph, blue circles indicate a Random Regular (RR) Graph and green squares indicate a Barab\'asi-Albert (BA) graph. In the right panel, the average degree is fixed $z = 5$ for ER and RR, and $z = 8$ for the BA. Predicted values by numerical integration of Eqs. \eqref{eq:PA} (solid lines) fit approximately the results for each topology. System size is fixed at $N = 3.2 \times 10^5$. }
\end{figure}

Linearization is also useful to determine the time dependence of the cascade process.  Assuming a small initial seed and rewriting the term $h(\phi)$ as  $ \sum_{l=0}^{\infty} K_l\,  \phi^l $, the linearized equation for the fraction of adopters $\rho(t)$ becomes
\begin{equation}
    \label{eq:linear_r}
    \frac{d \rho}{dt} \approx  \langle p_A \rangle\,  ( K_1 -1)\, \phi,
\end{equation}
where the coefficients $K_l$ are
\begin{equation}
    \label{eq:coef_rho}
    K_l = \sum_{k=l}^{\infty} \sum_{m=0}^{l} { k \choose l} \, {l \choose m} \, (-1)^{l+m}\,  p_k \, \theta\left( m/k - T \right).
\end{equation}

 A solution for the fraction of adopters $\rho(t)$ can be obtained from  Eqs. \eqref{eq:phi_general} and \ref{eq:linear_r}.  For the case of the Threshold model without aging, setting $\langle p_A \rangle = 1$,  the solution is an exponential cascade dynamics
\begin{equation}
    \rho(t) = \rho_0 \, e^{(C_1 - 1)\, t}.
\end{equation}
Therefore, the number of adopters $\rho (t)$ follows an exponential increase with exponent $\alpha(z,T)$:
\begin{equation}
    \label{eq:alpha}
    \alpha(z,T) = C_1 - 1 = \sum_{k=0}^{\lfloor 1/T \rfloor} \frac{k \, (k - 1)}{z}\, p_k - 1,
\end{equation}
where $C_1$ is computed from Eq. \ref{eq:coef_phi}. 

For both endogenous and exogenous aging, the same derivation is valid to determine the exponents $\delta(z,T)$ and $\gamma(z,T)$. For the case of endogenous aging ($\langle p_A \rangle = 1/(t+2)$), the fraction of adopters follows a power law dependence,
\begin{equation}
    \label{rho_endo}
    \rho(t) = \rho_0 \, \left( \frac{t+2}{2} \right)^{(C_1 - 1)}.
\end{equation}
The exponent reported for the power-law cascade dynamics $\delta(z,T)$ turns out to be, therefore, the same exponent as the one for the exponential behavior where there is no aging:  $\delta(z,T)= \alpha(z,T)$. Fig. \ref{fig:endo_exp} compares the prediction of Eq. \eqref{eq:alpha} with the results computed from numerical simulations. There is a good agreement for both Barab\'asi-Albert and Erd\"os-R\'enyi networks for all values of $T$ and $z$. For a random-regular graph, the predicted dependence, $\alpha(z) = z - 2$, is not a good approximation for large $z$. This is because the presence of small cycles increase importantly in a random-regular graph as the average degree $z$ grows \cite{wormald_1999} and the locally-tree assumption made for the derivation of the rate equations (Eq. \eqref{eq:AME_Threshold}) is not valid anymore. A different approach is necessary for clustered networks (as in Ref.\cite{Leah2022} for the Threshold model). In addition, endogenous aging with a general activation probability $p_A(j) = \frac{a}{j+2}$ allows tuning the cascade exponent with an additional control parameter $a$ $\left( \delta_a(z,T) = a \, \alpha(z,T) \right)$, what might be useful to fit real data.

\begin{figure}
\includegraphics[width=\columnwidth]{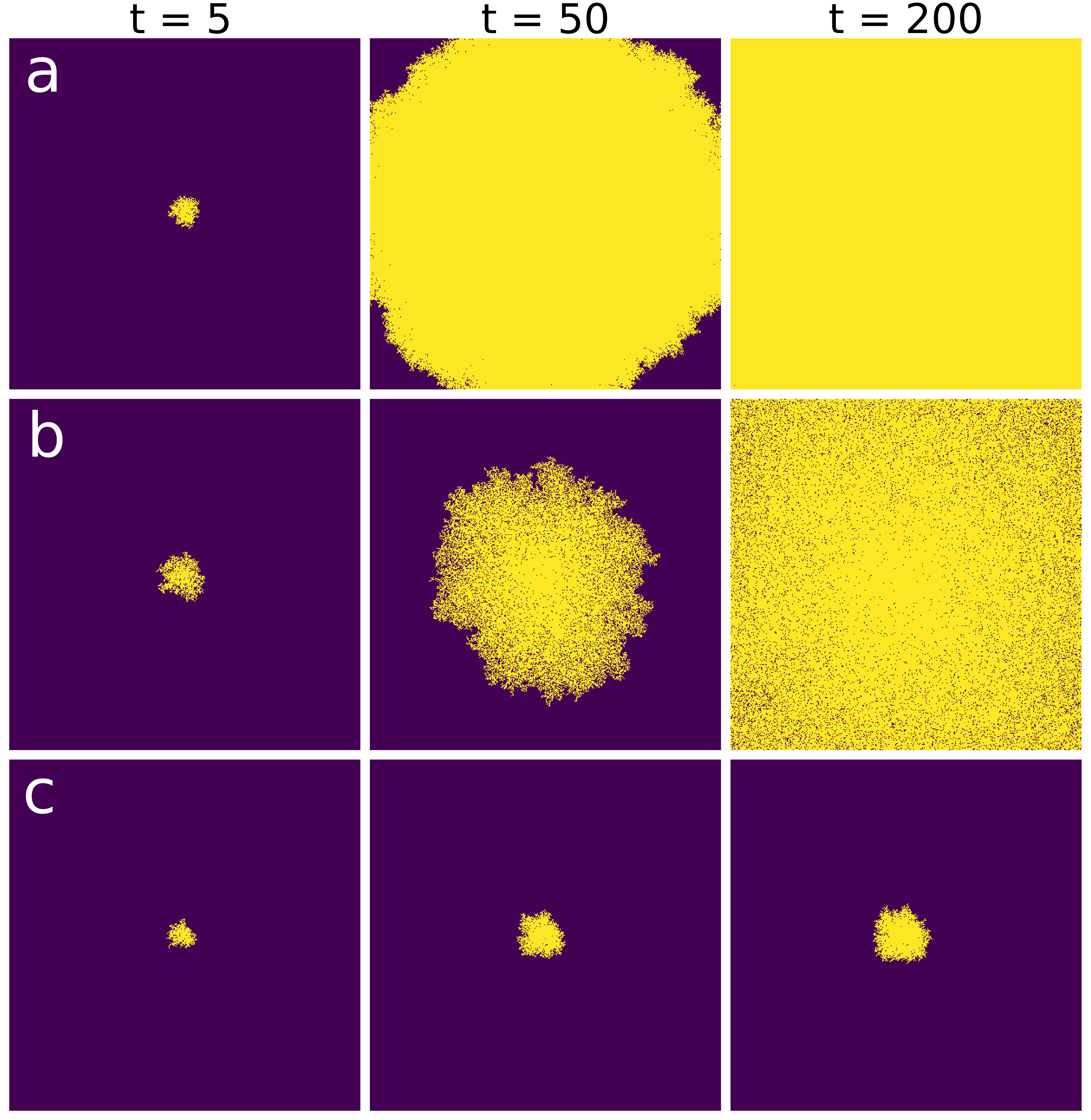}
\caption{\label{fig:evo_lat} Cascade spreading of the original Threshold model (a) and the versions with exogenous (b) and endogenous (c) aging on a Moore neighborhood lattice with size $N = L \times L$, $L = 405$. Yellow and purple nodes are adopters and non-adopters, respectively. Time increases from left to right. Initial seeds are selected favoring cascades: one agent and all her neighbors are set as adopters at the center of the system.}
\end{figure}

For exogenous aging, an analytical expresion for the exponent $\gamma(z,T)$ is not obtained following this methodology. Still, we can fit the exponent from the integrated solutions in Fig. \ref{fig:exo_endo_evo} (d-f). Fig.\ref{fig:exo_exp} shows the good comparison between the exponent calculated from the integration of the approximate equation and the one calculated from  numerical simulations. The dependence of $\gamma(z,T)$ with the parameters $z$ and $T$ is qualitatively similar to the dependence of  $\alpha(z,T)$ for the case of endogenous aging.

\section{\label{sec:Lattice} Lattice}

\begin{figure}
\includegraphics[width=\columnwidth]{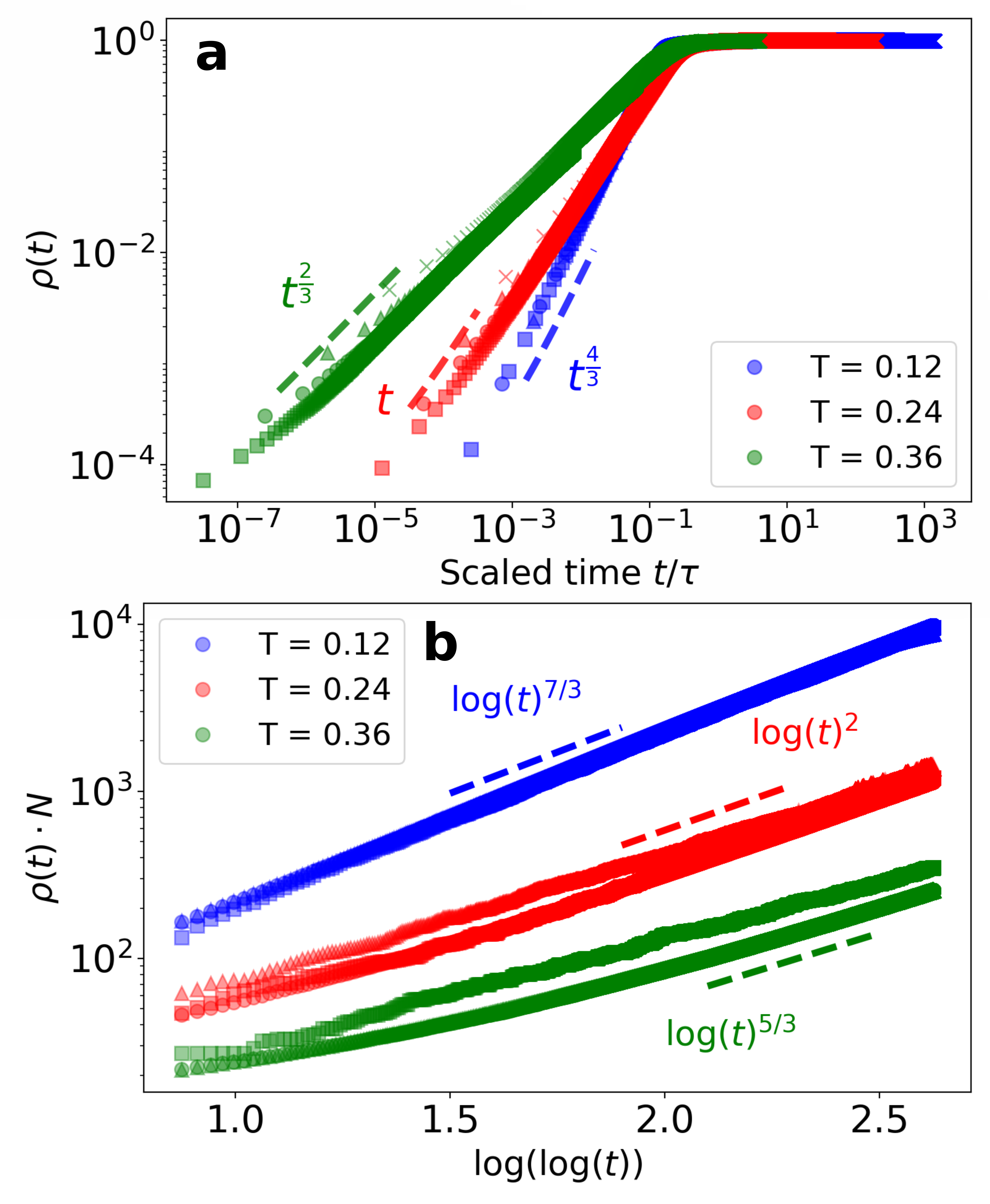}
\caption{\label{fig:lattice} Cascade dynamics of the Threshold model with exogenous (a) and endogenous (b) aging on a Moore neighborhood lattice. Different colors indicate different values of the threshold $T$. Different markers indicate the results of numerical simulations with different system size $N = L \times L$:  $L = 50$ (crosses), $100$ (triangles), $200$ (circles) and $400$ (squares). In (a), time is scaled according to size $\tau = L^{2 / \epsilon}$. Discontinuous solid lines indicate a power law behavior with exponent $ \epsilon = 4/3$ (blue), $1$ (red) and $2/3$ (green). In (b), the system sizes are not scaled due to the slow dynamics. Discontinuous solid lines indicate a power-logarithmic behavior, $\rho(t) \, N \sim \log(t)^{\nu} $, with exponent $ \nu = 7/3$ (blue), $2$ (red) and $5/3$ (green).}
\end{figure}

The Threshold model in a two-dimensional regular lattice with a Moore neighborhood (nearest and next nearest neighbors) is known to have a critical threshold $T_c = 3/8$ \cite{centola-2007}. Below this value, cascade dynamics follows a power-law increase in the density of adopters $\rho(t) \approx t^2$, which does not depend on the threshold value $T$. In Fig. \ref{fig:evo_lat}a, we show a typical realization of this model: From an initial seed, the adoption radius increases linearly with time until all agents adopt the technology.

When aging is considered, cascade dynamics become much slower and a dependence on $T$ appears. When the aging mechanism is exogenous, numerical simulations indicate a cascade dynamics following a power-law $\rho(t) \approx t^{\epsilon(T)}$. Qualitatively, we observe that while in the case without aging there was a soft interface between adopter and non-adopters, aging causes a strong roughening in the interface and the presence of non-adopters inside the bulk (see Fig. \ref{fig:evo_lat}b). In addition, the exponent values fitted from numerical simulations allow us to collapse curves for different system sizes (see Fig. \ref{fig:lattice}a). Due to finite size effects, the interface between adopters and non-adopters eventually reaches the borders of the system and the remaining non-adopters, in the bulk, will slowly adopt with the density of adopters following the functional shape $\rho(t) = 1- 1/(t+2)$.
 
Fig.\ref{fig:evo_lat}c shows the dynamics towards global adoption for endogenous aging. In comparison with the case of exogenous aging, we do not observe strong interface roughening between adopters and non-adopters and non-adopters do not exist in the bulk. Numerical simulations indicate a very slow increase of the density of adopters $\rho$, similar to a power-logarithmic growth  $\rho(t) \approx (\log(t))^{\nu}$, with a threshold dependent 
exponent $\nu(T)$  (Fig. \ref{fig:lattice}b).
 
Unfortunately, we were not able to find an analytical framework for the Threshold model in a regular lattice. Our general approximation used for complex networks assume a tree-like network, and it is not appropriate for this case. 

\section{\label{sec:Summary and Conclusions} Conclusions}

We have addressed in this work the role of aging in general models with binary-state agents interacting in a complex network. Temporal activity patterns are incorporated by means of a variable that represents the internal time of each agent. We have developed an approximate Master Equation for this general situation. In this framework, we have explicitly studied the effect of aging in the Threshold model as a paradigmatic example of Complex Contagion processes. Aging implies a lower probability to change state when the internal time increases. We have considered endogenous aging in which the internal time measure the persistence time in the same state, and exogenous aging in which the internal time measures the time since the last update attempt.


Our theoretical framework with some approximations to attain analytical results provide a good description of the results from numerical simulations for Erd\"os-R\'enyi, random-regular and Barab\'asi-Albert networks. For these three types of complex networks, we find that the cascade condition $T_c$ (critical value of the threshold parameter $T$ as a function of mean degree $z$ of the network) for the full spreading from an initial seed is not changed by the aging mechanisms. However, aging modifies in non-trivial ways the dynamics of the cascade process. The exponential growth with exponent $\alpha(z,T)$ of the density of adopters in the absence of aging becomes a power law with exponent $\delta(z,T)$ for endogenous aging, and a stretched exponential characterized by an exponent $\gamma(z,T)$ for exogenous aging. We have analysed the exponents dependence $\alpha(z,T)$, $\delta(z,T)$, $\gamma(z,T)$ and shown that $\alpha(z,T)=\delta(z,T)$.

Our general theoretical framework, based on the assumption of a tree-like network, is not appropriate for a regular lattice. In this case, we have been only able to run numerical simulations. Our results indicate that  exogenous aging gives rise to a adoption dynamics characterized by an increase in the roughness of the interface between adopters and non-adopters, by the presence of non-adopters in the bulk and by a power law growth of  the density of adopters with exponent $\epsilon (T)$, while in the absence of aging $\epsilon = 2$ independently of $T$. Endogenous aging, on the other hand, produces a very slow (logarithmic like) spreading dynamics.

This work highlights the importance of non-Markovian dynamics in general  binary-state dynamics and, specifically, in the Threshold model of complex contagion. The theoretical framework presented here gives a basis for further investigations of the memory effects and non-Markovian dynamics in networks, and in particular for  binary-state models with aging. Still, a number of theoretical developments remain open for future work, such as the consideration of stochastic finite size effects \cite{peralta-2020B}. Also, proper approximations need to be developed to account for some of our numerical results for random-regular networks with high degree, as well as for high clustering , degree-degree correlations networks and for regular lattices, including continuous field equations for this latter case. 

\vspace{-0.5 cm}

\begin{acknowledgments}

Partial funding is acknowledged from the project PACSS (RTI2018-093732-B-C21, RTI2018-093732-B-C22) of the MCIN/AEI/10.13039/501100011033/ and by EU through FEDER funds (A way to make Europe), and also from the Maria de Maeztu program MDM-2017-0711 of the MCIN/AEI/10.13039/501100011033.

\end{acknowledgments}

\vspace{-0.5 cm}

\appendix

\section{\label{app:DERIVATION OF MASTER EQUATION WITH AGING} DERIVATION OF A GENERAL MASTER EQUATION FOR  BINARY-STATE MODELS WITH AGING IN COMPLEX NETWORKS}

\begin{figure}
\includegraphics[width=\columnwidth]{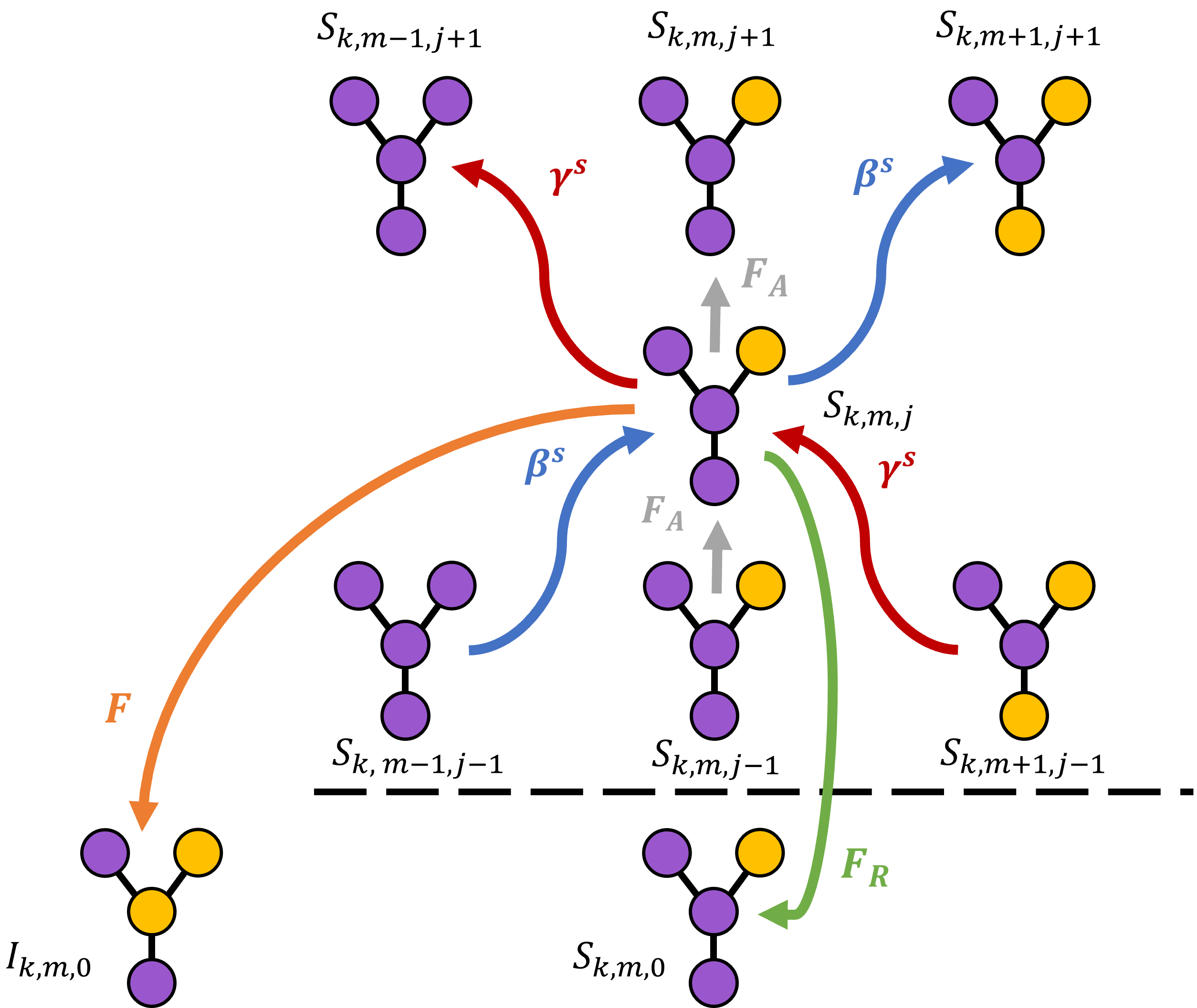}
\caption{\label{fig:ame_plot1} Schematic representation of the transitions to or from the set $s_{k,m,j}$ ($j > 0)$. We show the central node with some neighbors for different values $m$ and $j$. Purple nodes are susceptible or non-adopters or spin-down, and yellow are infected or adopters or spin-up.}
\end{figure}

We consider  binary-state dynamics on static, undirected, connected networks in the limit of local tree-like structure, following closely the approach used in Ref. \cite{gleeson-2013} for binary-state dynamics in complex networks. The new ingredient  is to consider the nodes with different age as different sets, what allows us to treat as Markovian the memory effects introduced by aging \cite{peralta-2020C,peralta-2020A}. We define $s_{k,m,j} (t)$ ($i_{k,m,j} (t)$) as the fraction of nodes that are susceptible (infected) and have degree $k$, $m$ infected neighbors and age $j$ at time $t$. The initial condition is set such that all agents have age $j = 0$ and there is a randomly chosen fraction $\rho_0$ of nodes infected:
\begin{align} \label{initial_condition} 
\textrm{For } j > 0 & \quad    s_{k,m,j} (0) = 0  \quad   i_{k,m,j} (0) = 0, \nonumber\\
\\
\textrm{For } j = 0 & \quad    s_{k,m,0} (0) = (1 -  \rho_0)\, B_{k,m}[\rho_0] \nonumber \\
                    & \quad    i_{k,m,0} (0) = \rho_0\, B_{k,m}[\rho_0], \nonumber
\end{align}
where $B_{k,m}[\rho_0]$ is the binomial distribution with $k$ attempts, $m$ successes and $\rho_0$ is the initial fraction of infected agents that as the probability of success of the binomial. Now, we examine how $s_{k,m,j}$ changes in a time step. We consider separately the case $j = 0$ since its evolution is different from $j > 0$. See Fig. \ref{fig:ame_plot1} for a schematic representation of transitions involving $s_{k,m,j}$. This is the way to reach the expressions of Eq. \eqref{eq:pre_AME}:
\begin{widetext}
\begin{align} \label{eq:pre_AME}
    s_{k,m,j} (t + dt) = & \, s_{k,m,j}(t) - F (k,m,j)\, s_{k,m,j}\, dt - F_{R} (k,m,j)\, s_{k,m,j} \, dt - F_{A} (k,m,j) \, s_{k,m,j} \, dt \nonumber \\
    & + F_{A} (k,m,j-1)\,  s_{k,m,j-1} \, dt - \omega (s_{k,m,j} \to s_{k,m+1,j+1}) \, s_{k,m,j}\, dt  \nonumber \\
    & - \omega (s_{k,m,j} \to s_{k,m-1,j+1})\,  s_{k,m,j} \, dt + \omega (s_{k,m+1,j-1} \to s_{k,m,j}) \, s_{k,m+1,j-1} \, dt \nonumber \\
    & + \omega (s_{k,m-1,j-1} \to s_{k,m-1,j-1}) \, s_{k,m-1,j-1}\,  dt, \\
    s_{k,m,0} (t + dt) = &\,  s_{k,m,0}(t) - F (k,m,0) \, s_{k,m,0} \, dt + \sum_{l = 0} R (k,m,l)\,  i_{k,m,l} \, dt + \sum_{l = 1} F_{R} (k,m,l)\,  s_{k,m,l}\,  dt   \nonumber\\
    & - F_{A} (k,m,0)\,  s_{k,m,0}\,  dt - \omega (s_{k,m,0} \to s_{k,m+1,1}) \, s_{k,m,0}\,  dt - \omega (s_{k,m,0} \to s_{k,m-1,1})\,  s_{k,m,0} \, dt .\nonumber
\end{align}
\end{widetext}

Similar equations can be found considering changes in $i_{k,m,j}$. In these equations, the transition probabilities (described in detail in section \ref{subsec:Approximate master equation and solutions}) allow agents to change state ($F$ and $R$), reset internal time ($j \to 0$) ($F_R$ and $R_R$ and age ($j \to j + 1$) ($F_A$ and $R_A$). Notice that we have considered no transition increasing (or decreasing) the number of infected neighbors $m$, keeping constant the age $j$. This is because the age $j$ is defined as the time spent in the current state (or since a reset). Therefore, if a node remains susceptible and the number of infected neighbors changes ($m \to m \pm 1$), the age of the node must increase ($j \to j + 1$). To determine the rate of these events, we use the same assumption as in Ref. \cite{gleeson-2013}: we assume that the number of S-S edges change to S-I edges at a time-dependent rate $\beta^s$. Therefore, the transition rates are:
\begin{align} \label{rate_beta_s}
&  \omega (s_{k,m,j} \to s_{k,m+1,j+1}) = (k - m) \, \beta^s, \nonumber \\
\\
& \omega (s_{k,m-1,j-1} \to s_{k,m,j}) = (k - m + 1)\, \beta^s . \nonumber 
\end{align}

To determine the rate $\beta^s$, we count the change of S-S edges that change to S-I in a time step. This change is produced by a neighbor changing state from susceptible to infected. Thus, we can extract this information from the infection probability $F (k,m,j) $:
\begin{equation}
    \label{beta_s}
    \beta^s = \frac{\sum_j \sum_k p_k \sum_{m = 0}^{k} (k - m)\, F (k,m,j) \, s_{k,m,j}}{\sum_j \sum_k p_k \sum_{m = 0}^{k} (k - m) \, s_{k,m,j}}.
\end{equation}
A similar approximation is used to determine the transition rates at which S-I edges change to S-S edges. We write:
\begin{align} \label{rate_gamma_s}
&  \omega (s_{k,m,j} \to s_{k,m-1,j+1}) = m\, \gamma^{s}, \nonumber \\
\\
& \omega (s_{k,m+1,j-1} \to s_{k,m,j}) = (m + 1)\, \gamma^{s} , \nonumber
\end{align}
where the rate $\gamma^s$ is computed using the recovery probability $R (k,m,j)$:
\begin{equation}
    \label{gamma_s}
    \gamma^s = \frac{\sum_j \sum_k p_k \sum_{m = 0}^{k} (k - m)\, R (k,m,j)\, i_{k,m,j}}{\sum_j \sum_k p_k \sum_{m = 0}^{k} (k - m)\,  i_{k,m,j}}.
\end{equation}

For standard models, one natural assumption is to consider the probability to age as the probability of neither changing state nor resetting:
\begin{align} \label{cond_1}
&  F (k,m,j) + F_{A}(k,m,j) + F_{R}(k,m,j) = 1, \nonumber\\
\\
&  R (k,m,j) + R_{A}(k,m,j) + R_{R}(k,m,j) = 1. \nonumber
\end{align}
With this condition, taking the limit $dt \to 0$ of Eq. \eqref{eq:pre_AME} we obtain the approximate master equation (AME) for the evolution of the different sets $s_{k,m,j}$, $s_{k,m,0}$ $i_{k,m,j}$ and $i_{k,m,0}$:
\begin{align}
\label{eq:AME}
    \frac{d s_{k,m,j}}{dt} = & \, - s_{k,m,j} - (k - m)\, \beta^s\,  s_{k,m,j} - m \, \gamma^s\, s_{k,m,j} \nonumber \\
    & + (k-m+1)\, \beta^s \,   s_{k,m-1,j-1} \nonumber\\
    & + (m+1)\, \gamma^s \,  s_{k,m+1,j-1}\nonumber\\
    & + F_{A}(k,m,j-1)\,  s_{k,m,j-1},  \nonumber\\
    \frac{d s_{k,m,0}}{dt}  = & \, - s_{k,m,0} - (k - m) \, \beta^s\,  s_{k,m,0} - m\, \gamma^s \,  s_{k,m,0} \nonumber \\
    & + \sum_{l = 0} R (k,m,l)\,  i_{k,m,l} + \sum_{l = 0} F_{R}(k,m,l)\, s_{k,m,l}, \nonumber \\
     \frac{d i_{k,m,j}}{dt}  = & \, - i_{k,m,j} - (k - m)\, \beta^i\,  i_{k,m,j} - m\, \gamma^i \,   i_{k,m,j} \nonumber\\
    & + (k-m+1)\, \beta^i\,    i_{k,m-1,j-1} \nonumber\\
    & + (m+1)\, \gamma^i \,   i_{k,m+1,j-1} \\
    & + R_{A}(k,m,j-1) \, i_{k,m,j-1},  \nonumber\\
     \frac{d i_{k,m,0}}{dt}  = & \,  - i_{k,m,0} - (k - m) \, \beta^i\,  i_{k,m,0} - m\,  \gamma^i \,  i_{k,m,0} \nonumber\\
    & + \sum_{l = 0} F(k,m,l) \, s_{k,m,l} + \sum_{l = 0} R_{R}(k,m,l) \, i_{k,m,l}, \nonumber
\end{align}
where $\beta^i$ and $\gamma^i$ are similar rates as $\beta^s$ (Eq. \ref{beta_s}) and $\gamma^s$ (Eq. \eqref{gamma_s}), exchanging terms $s_{k,m,j}$ by $i_{k,m,j}$ and vice versa. These equations define a closed set of deterministic differential equations that can be solved numerically using standard computational methods for any complex network and any model aging via the infection/recovery, reset and aging probabilities (a general script in Julia is available in author's GitHub repository \cite{link_git}).

The model is introduced via the transition probabilities ($F, R, F_A, R_A, F_R, R_R$), which may depend on the degree $k$, the number of infected neighbors $m$ and the time spent in the actual state (or since a reset) $j$. For the Threshold model with aging, dynamics are monotonic and there are no age dynamics once the agent is infected $R (k,m,j) = R_A (k,m,j) = R_R(k,m,j) = 0$. Therefore, the equations for $s_{k,m,0}$  decouples from the equations for the variables $i_{k,m,j}$, reducing Eq.\eqref{eq:AME} to:
\begin{align}
\label{eq:AME_Threshold_AP}
    \frac{d s_{k,m,j}}{dt}  = & \, - s_{k,m,j} - (k - m)\, \beta^s \,   s_{k,m,j} \nonumber\\
    & + (k-m+1) \, \beta^s \,  s_{k,m-1,j-1}  \nonumber\\
    & + F_{A} (k,m,j-1)\,  s_{k,m,j-1},  \\
    \frac{d s_{k,m,0}}{dt}  = & \,  - s_{k,m,0} - (k - m)\, \beta^s \,  s_{k,m,0} \nonumber \\
    & + \sum_{l = 0} F_{R}(k,m,l) \, s_{k,m,l} . \nonumber
\end{align}

\end{document}